\documentclass[twocolumn,prl,showpacs,amsmath,amssymb,floatfix]{revtex4}
\usepackage{graphicx}
\usepackage{dcolumn}
\usepackage{bm}

\begin{document}

\title{Instability in a Network Coevolving with a Particle System}

\author{Sang-Woo Kim and Jae Dong Noh}
\affiliation{Department of Physics, University of Seoul,
  Seoul 130-743, Korea}

\date{\today}
\begin{abstract}
We study a coupled dynamics of a network and a particle system. 
Particles of density $\rho$ diffuse freely along edges, 
each of which is rewired at a rate given by a decreasing function
of particle flux. We find that the coupled dynamics 
leads to an instability toward the formation of hubs and that 
there is a dynamic phase transition at a threshold particle
density $\rho_c$. In the low density phase, the network evolves into a
star-shaped one with the maximum degree growing linearly 
in time. In the high density phase, the network exhibits a 
fat-tailed degree distribution and an interesting dynamic scaling behavior.
We present an analytic theory explaining mechanism for the 
instability and a scaling theory for the dynamic scaling behavior.

\end{abstract}
\maketitle

For the past decade growing interests have been paid to complex networks.
They are ubiquitous in nature and display intriguing properties which
have not been observed in periodic lattices or random networks.
The work of Ref.~\cite{Watts98} triggered extensive and intensive studies on 
structure and dynamics of complex 
networks~\cite{Albert02,Dorogovtsev02,Newman03}. 
Dynamics and cooperative phenomena in various systems 
defined on networks have also attracted a lot of 
attention~\cite{Boccaletti06,Dorogovtsev07}. 
Most studies so far have considered
dynamics {\em of} networks or dynamics {\em on} networks separately.
The aim of the present work is to investigate emerging 
structure of a network coevolving with a dynamical system.

Properties of dynamical systems or models for cooperative phenomena 
are strongly affected by underlying network structure. For instance, 
the study on
random walks~\cite{Noh04} shows that the density of diffusing 
particles at nodes is strictly proportional to the degree of nodes and that
the mean first passage time is determined by the network structure through
the so-called random walk centrality.  
Importance of underlying network structure is also shown in the study of
critical phenomena~\cite{Dorogovtsev07}, condensation~\cite{Noh05}, opinion
dynamics~\cite{Sood05}, and so on.

Just as network structure affects dynamics on it, the former 
may also be influenced from the latter. 
The synaptic plasticity is an example of such phenomena~\cite{Lomo03}. 
In neural networks, bio-chemical signals are
transmitted from neuron to neuron through synaptic links. At the same time, 
the strength of synapses can be enhanced or suppressed depending on synaptic
activities. It is called the synaptic plasticity, which may result in
deformation of neural networks.

When structure and dynamics are coupled, the interplay
between them will drive a network to evolve in a self-organized way.
It is challenging to study the emerging property of such a network.
We will show that the interplay 
can lead to an instability toward the formation of hubs.
There are a few recent works on coevolutionary dynamics of complex networks.
Network dynamics combined with a game theoretical model was studied 
in Refs.~\cite{Ebel02,Zimmermann04},
and that combined with a voter model type opinion
dynamics was studied in Refs.~\cite{Gil06,Holme06}. 
However, the dynamic instability was not observed in those studies.

We study a minimal model which consists of a network and diffusing particles.
A network is undirected and consists of $N$ nodes. Each edge
$e=(i,j)$ between nodes $i$ and $j$ is assigned to a positive weight 
$w_e$. There are particles of density $\rho$ distributed over 
nodes. We adopt the following dynamic rule:
(i) All particles hop to their neighboring nodes randomly and 
independently. (ii) If a particle hops
from node $i$ to $j$, the weight of all edges attached to $j$ is increased
by unity. (iii) After the hopping of all particles, each edge $e$ is rewired
with the probability $1/w_e$. The weight of rewired edges is 
set to unity. The time is increased by unity after those processes.

The diffusion (i) mimics a transport taking place on a network.
For simplicity, the particles are taken to be non-interacting. 
According to (ii), the weight $w_e$ of an edge $e=(i,j)$ established at
time $t_e$ is given by 
\begin{equation}\label{w_e}
w_e(t) = 1+\sum_{t'=t_e}^t ( n_i(t')+n_j(t')).
\end{equation}
Here, $n_i(t')$ denotes the number of particles visiting node $i$ 
at time $t'$. 
The more an edge contributes to a transport the more robust 
it is~\cite{comment0}.  Less important edges
are weeded out and replaced by new ones in the process (iii). 

We start with a random network with $N$ nodes and mean degree $\langle
k\rangle$ over which particles of density $\rho$ are distributed randomly. 
The weight of all edges are set to unity. 
Then we measure the degree $k_{max}$ of the
node having the largest degree and the degree distribution $P_{deg.}(k)$,
which are averaged over $N_S$ samples. The mean degree is fixed to $\langle
k\rangle=4$ and $N_S=10^3$ in numerical studies.

\begin{figure}[t]
\includegraphics*[width=\columnwidth]{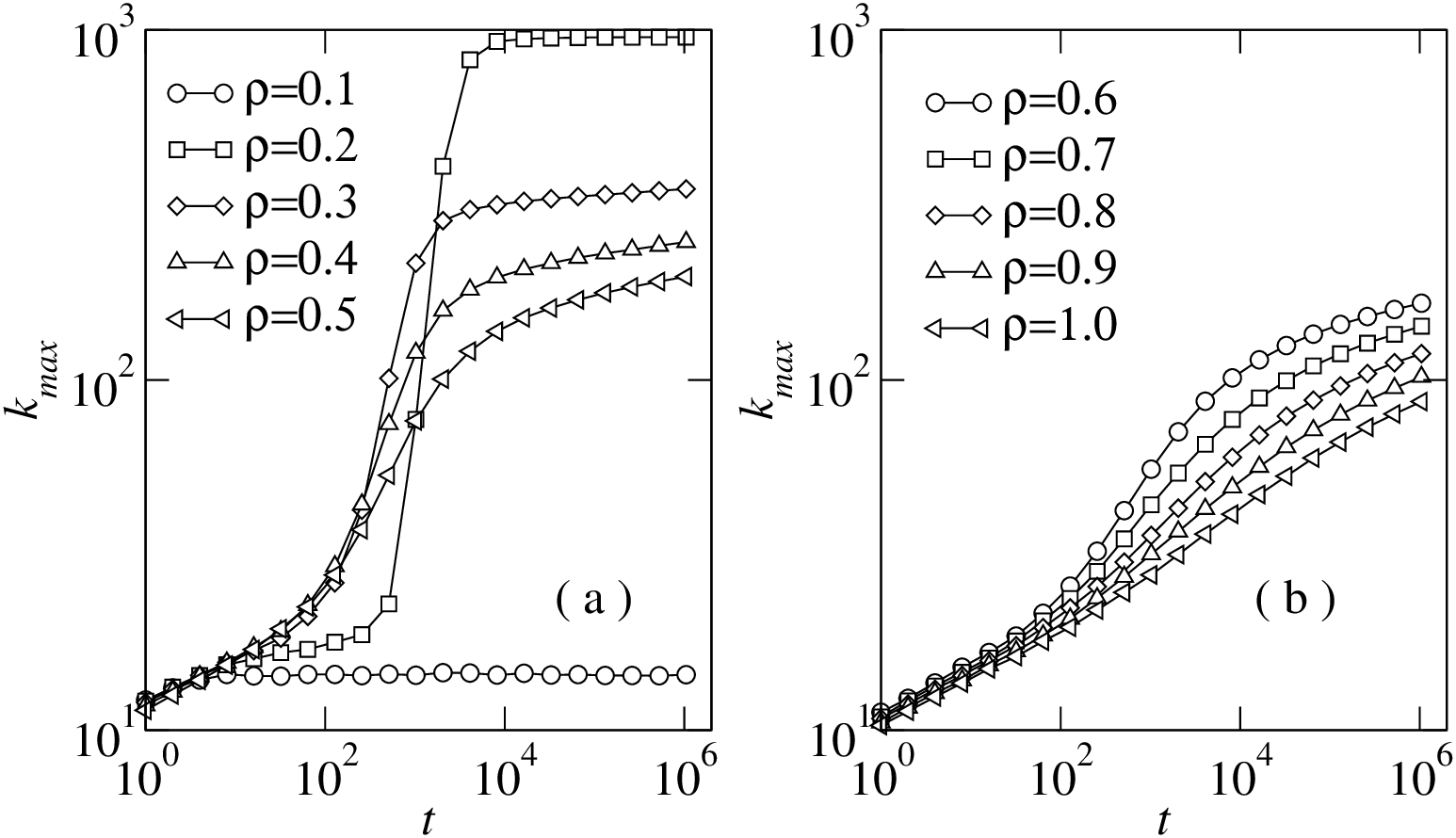}
\caption{Time evolution of $k_{max}$ in networks with $N=10^3$.}\label{fig1}
\end{figure}

Figure~\ref{fig1} shows the numerical data for $k_{max}$ with $N=1000$. 
One finds that
$k_{max}$ increases in time exceeding the value $k_{max} =
\mathcal{O}(\ln{N})$ 
which one would expect in random networks at all values of
$\rho$ except $0.1$. This suggests that
there exists a dynamic instability toward the formation of hubs. 
Initially all edges have low weights and they are rewired randomly at a
constant rate. Suppose that a node $i$ happens to be linked with more edges 
than others due to a statistical fluctuation. Then it will be visited by more
particles since diffusing particles tend to be attracted toward higher
degree nodes~\cite{Noh04}. This will strengthen the edges emanating from $i$, 
and the node $i$ will have more chance to increase its degree.
This feedback may be a possible mechanism for the instability. 
This idea will be elaborated in detail later.

\begin{figure}[t]
\includegraphics*[width=\columnwidth]{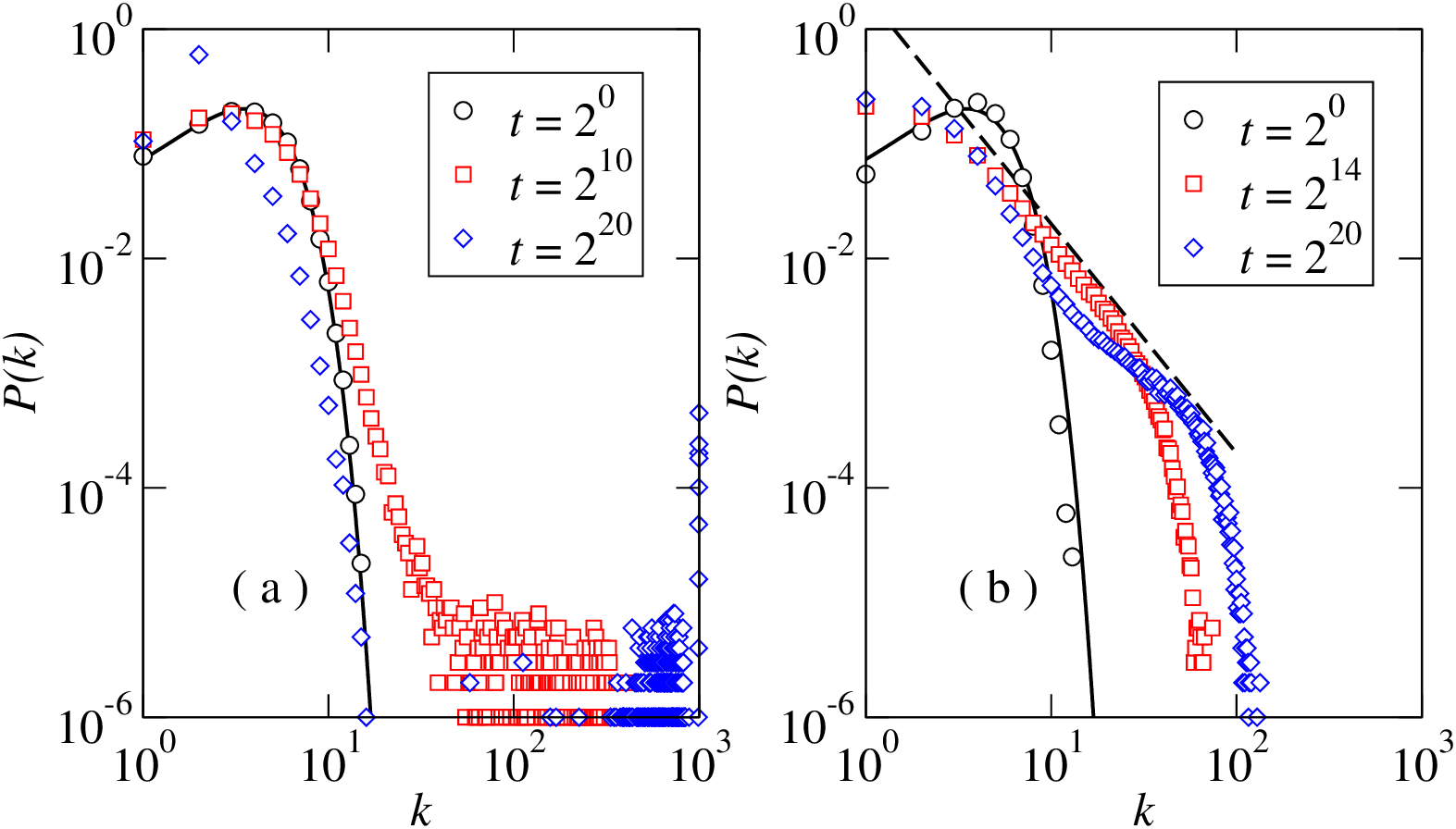}
\caption{Time evolution of the degree distribution of the networks with
$N=10^3$ and  with (a) $\rho=0.2$ and (b) $\rho=1.0$. The solid curves
represent the Poisson distribution $P_{deg.}(k) = e^{-\langle
k\rangle} \langle k \rangle^k / k!$ with $\langle k \rangle = 4$. The dashed
line in (b) has the slope $-2$.}
\label{fig2}
\end{figure}

The numerical data in Fig.~\ref{fig1} also
suggest that there is a dynamic phase transition at $\rho=\rho_c \simeq 0.6$. 
The threshold will be estimated from a scaling theory which will
be presented later.
When $\rho$ is small~(see Fig.~\ref{fig1}(a)), $k_{max}$
remains almost constant up to a certain time scale $\tau$. Then it 
grows ballistically as $k_{max} \sim t$
until it reaches the limiting value $k_{max} \simeq N$. 
We will call $\tau$ the instability time. 
More detailed information is obtained from the 
degree distribution presented in Fig.~\ref{fig2}(a). 
It follows the Poisson distribution for $t\ll\tau$, 
which indicates that all nodes are statistically equivalent and edges are 
being rewired randomly. At $t \simeq \tau$, a
hub emerges spontaneously developing a peak in the degree
distribution. The hub grows until it is connected to almost all
other nodes. Finally there is an isolated peak in the degree distribution
and the network becomes star-like.

The system exhibits distinct behaviors when $\rho$ is large~(see
Fig.~\ref{fig1}(b)).
The instability sets in immediately and then
$k_{max}$ increases sublinearly in time, whose time dependence has not been
characterized yet.  
The numerical data in Fig.~\ref{fig2}(b) show 
that the degree distribution remains continuous and keeps broadening.
These behaviors allow us to interpret that hubs emerge simultaneously and
compete with each others to grow into larger ones. 
During the growth, the degree distribution can be fitted into the
power-law form as $P_{deg.}(k) \sim k^{-\gamma}$ with $\gamma\simeq 2.0$.
The power-law degree distribution persists for a long time, but 
is not a stationary one. 
The numerical data show that there appears a dip in the intermediate $k$
regime. It suggests that a single hub will dominate 
and the network will become star-like eventually, which we could 
not observe numerically up to $t=\mathcal{O}(10^7)$ though.

We present a phenomenological theory that explains mechanism for 
the instability. 
On a non-evolving complex networks, a diffusing particle 
relaxes quickly to the stationary state 
in which the visiting frequency to a node is strictly proportional 
to its degree~\cite{Noh04}. 
Using this property, we assume that the diffusing
particles remain in the quasi-stationary state to a given network at each
moment. The quasi-stationarity assumption allows us to make the 
approximation $n_i(t) \simeq \rho k_i(t)/\langle k \rangle$ in Eq.~(\ref{w_e}),
with which we can eliminate the particles degrees of freedom.

In order to describe the onset of the instability, it suffices to consider
an effective dynamics of a single node $I$ and its degree $K$.
Before the onset, all edges in the network are 
rewired randomly at a constant rate. 
So we can assume that $K$ is increased~($K\rightarrow K+1$) at 
each time step with a suitable choice of time unit.
The weight $w_\alpha$ of each edge $\alpha=1,\cdots,K$ is set to unity 
when it is attached to $I$, and then increased by the amount of $\Delta
w_\alpha = \lambda K(t)$~\cite{comment1} 
at time step $t$ according to the quasi-stationarity assumption. 
The constant factor $\lambda$ should be an increasing function of $\rho$, 
whose explicit form is not necessary. 
So, the weight of an edge $\alpha$ having been 
attached to $I$ since time $t_\alpha$, is given by
\begin{equation}\label{w_alpha}
w_\alpha(t) = 1 + \lambda \sum_{t'=t_\alpha}^t K(t') \ .
\end{equation}
The degree $K$ decreases when an edge $\alpha$ is rewired 
with the probability $1/w_\alpha$. Combining those processes, we
can write down the rate equation for the time evolution of the mean value of
the degree:
\begin{equation}
\Delta K \equiv K(t+1) - K(t) = f_{in} - f_{out}, 
\end{equation}
where the incoming flux is given by $f_{in}=1$ and the outgoing flux is
given by $f_{out} = \sum_{\alpha=1}^{K(t)} 1 / w_\alpha(t)$. 
Since the weight given in Eq.~(\ref{w_alpha}) 
has history dependence, the effective dynamics for $K(t)$ is non-Markovian.
Note that the effective dynamics of a single node is valid only when edges
among other nodes are rewired at a constant rate~($f_{in}=1$). 

The resulting single node dynamics is analogous to that of a queueing
model~\cite{Gross98}. In that context, the node $I$, the edges, 
and the degree $K$ correspond to a queue, data packets, and the queue size, 
respectively.  
Hereafter, we will adopt the terminology of a queueing model for the single 
node problem.
Such a correspondence between
network dynamics and particles dynamics was also
considered in the context of a zero-range process~\cite{Angel05}.

\begin{figure}[t]
\includegraphics*[width=\columnwidth]{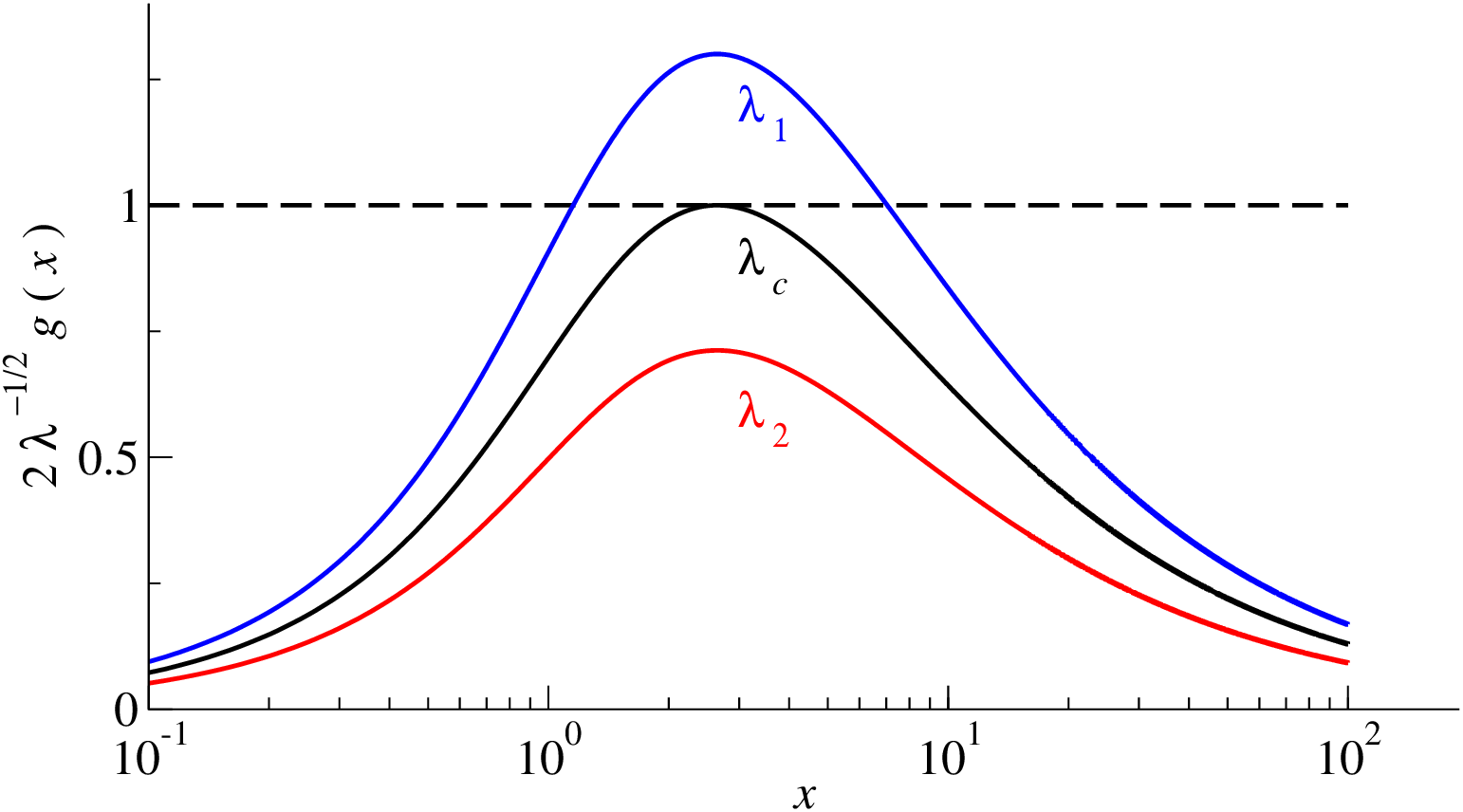}
\caption{Graph of $\frac{2}{\sqrt{\lambda}} g(x)$ with
$\lambda_1 < \lambda_c \simeq 2.03< \lambda_2$. }\label{fig3}
\end{figure}

Suppose that there are $K$ packets in the queue at time $t$. We denote by
$t_\alpha$~($\alpha=1,\cdots,K$) the time at which a packet $\alpha$ 
entered the queue. Labeling the packets in such a way that 
$\alpha<\alpha'$ implies $t_\alpha < t_{\alpha'}$, 
one has the inequality $K(t_\alpha)\geq \alpha$. It yields 
\begin{equation}\label{w_alpha_in}
w_\alpha \geq 1 + \lambda \sum_{\alpha'=\alpha}^K \alpha' ,
\end{equation}
which imposes an upper bound on the outgoing flux $f_{out} \leq
F_{out}(K,\lambda)$. A straightforward algebra shows that
\begin{equation}\label{F_out}
F_{out}(K,\lambda) = \frac{2}{\sqrt{\lambda}} g(\sqrt{\lambda} K) ,
\end{equation}
where 
\begin{equation}\label{g(x)}
g(x) = \frac{1}{\sqrt{x^2+2}} \ln 
\left(\frac{ x^2+2+ x \sqrt{x^2+2}}{2}\right)  .
\end{equation}
The shape of this function is drawn in Fig.~\ref{fig3}.
It attains the maximum value $g_c \simeq 0.712$ at $x=x_c \simeq 2.64$.

Note that the function $g(x)$ converges to zero as 
$g(x)\simeq \frac{x}{2}$ 
for $|x| \ll x_c$ and $g(x) \simeq \frac{2\ln x}{x}$ for $x\gg x_c$. 
Consequently $f_{out}~(<F_{out})$ decays to zero at sufficiently 
large values of $K$
at any nonzero value of $\lambda$ while $f_{in}=1$. 
It implies that the queue size will
diverge in the long time limit. However, dynamic features may be different
depending on the value of $\lambda$:
(i) When $\lambda > \lambda_c \equiv (2g_c)^2
\simeq 2.03$, $\Delta K = f_{in}-f_{out}>0$ for all values of $K$. Hence  
the queue size $K(t)$ grows immediately and asymptotically linearly 
in time.
(ii) When $\lambda < \lambda_c$, there may be a dynamic barrier in 
an interval $K_1< K < K_2$ where $\Delta K < 0$. 
In that case the queue can be trapped to an attractor at $K(t) = K_1$. 
It, however, cannot stay there permanently because the queue can escape from
the barrier due to a statistical fluctuation in 
a characteristic time scale $\tau$.
For $t> \tau$, the queue size $K(t)$ will grow linearly in time asymptotically. 

For $\lambda < \lambda_c$, we can estimate the time scale $\tau$ roughly.
As a crude approximation, we regard Eq.~(\ref{w_alpha_in}) 
as an equality so that the 
result $\tau'$ obtained thus will provide a lower bound for $\tau$.
The queue size increases by unity at each time step 
if no packet escapes from the queue. It happens
with the probability $P_{no}(K,\lambda) 
= \prod_{\alpha=1}^K ( 1 - 1/w_\alpha)$.  For large $K$  it is 
approximated as $P_{no}\sim \exp(-\sum_{\alpha} 1/w_\alpha)= \exp(
-F_{out}(K,\lambda))$. Thus, we can estimate the probability to overcome 
the dynamic barrier at $K_1<K<K_2$ as
$P_{esc}(\lambda) = \prod_{K=K_1}^{K_2}
\left[\exp(-F_{out}(K,\lambda))\right]$
and the time scale as $\tau'=1/P_{esc.}(\lambda)$. 
Using Eqs.~(\ref{F_out}) and (\ref{g(x)}), 
we obtain that
\begin{equation}\label{tau'}
\tau' \sim \exp\left( - \frac{a (\ln \lambda)^2}{\lambda}\right)
\end{equation}
with a constant $a$. 

Using the knowledge from the effective single node dynamics, 
one can understand the dynamic property of the original model. 
We first consider the small $\rho$ case (corresponding to the case with 
$\lambda<\lambda_c$). Initially all nodes are trapped into the dynamic
barrier. That is to say, all edges are rewired randomly and
the degree of all nodes is fluctuating around the mean value 
$\langle k \rangle$. In the meanwhile a certain node may escape from the 
barrier in the instability time $\tau$ acquiring more and more edges. 
Once it happens, the number of particles available to all other nodes 
decreases, which leaves them into a deeper barrier. Consequently, the
network will become star-like eventually. 
This is consistent with the numerical observation presented 
in Figs.~\ref{fig1} and \ref{fig2}. 
A rough estimate of the instability time $\tau$ is given by 
Eq.~(\ref{tau'}) with $\lambda$ replaced by $\rho$. It increases very
rapidly as $\rho$ decreases. It explains the reason why we 
could not observe the instability at $\rho=0.1$ numerically. 

When $\rho$ is large~(corresponding to the case with
$\lambda>\lambda_c$), the single node picture predicts that 
the degrees of all nodes increase simultaneously since there is no dynamic 
barrier hindering growth. 
However, the simultaneous growth will give rise to competition
among nodes. One cannot apply the independent single node picture 
any more to the network dynamics.

The quasi-stationarity condition for diffusing particles is still acceptable
since the edge rewiring dynamics becomes slower under the competition. 
So, the weight $w_e$ of an edge $e$ will increase linearly in
time as $w_e\simeq c\rho t$ with a degree-dependent constant $c$ until it is
rewired.  Its rewiring dynamics is determined by the survival
probability $P_s(t)$ which is defined as the probability that the edge has
remained unrewired for $t$ time steps. Up to a leading order, it is given by
\begin{equation}
P_{s}(t) = \prod_{t'=1}^t \left( 1 - \frac{1}{c\rho t}\right) 
\simeq t^{-1/c\rho} \ .
\end{equation}

The power-law scaling of the survival probability suggests a scale invariant
network dynamics in the large $\rho$ regime.
Suppose that two networks with particle density 
$\rho$ and $\rho_0$ are of a similar shape at time scale $t$
and $t_0$, respectively. 
The similarity can be preserved during evolution if the survival
probabilities of the corresponding edges are the same. Therefore, we make
the scale invariance ansatz that
a network with the particle density $\rho$ at time $t$ and that 
with the density $\rho_0$ at time $t_0$
are equivalent provided that
\begin{equation}
t = t_0^{\rho/\rho_0} .
\end{equation}

\begin{figure}[t]
\includegraphics*[width=\columnwidth]{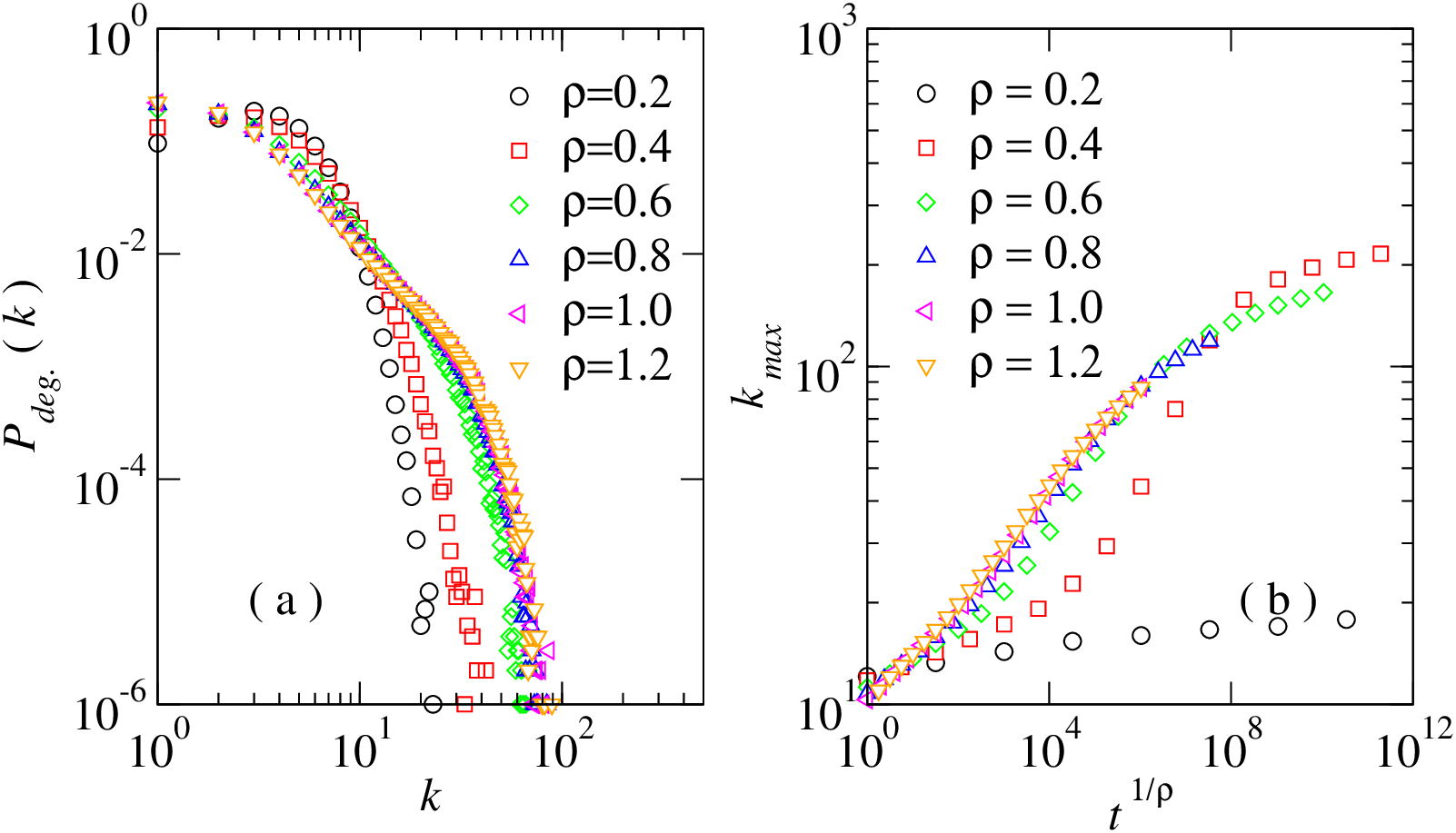}
\caption{(a) $P_{deg.}(k)$ for networks of $N=10^3$ nodes
at several values of $\rho$ and $t$ with fixed $t^{1/\rho}=2^{15}$.
(b) $k_{max}$ versus $t^{1/\rho}$.}\label{fig4}
\end{figure}
We examine the validity of the scale invariance numerically. It predicts
that the degree distribution $P_{deg.}(k)$ of networks with different values
of $\rho$ and $t$ should be the same if
$t^{1/\rho}$ is the same. In Fig.~\ref{fig4}(a) we present the numerical
data for $P_{deg.}(k)$ at several values of $\rho$ and $t$ with
fixed $t^{1/\rho}=2^{15}$. The data collapse well onto a single curve
for $\rho \gtrsim 0.6$. 
The scale invariance ansatz also predicts that the maximum
degree $k_{max}$ depends only the scaling variable  $t^{1/\rho}$. In
Fig.~\ref{fig4}(b), we plot the data for $k_{max}$ presented in
Fig.~\ref{fig1} with respect to the scaling variable $t^{1/\rho}$. The data
also collapse reasonably well onto a single curve for $\rho \gtrsim 0.6$.
From these analyses, we conclude that the dynamics displays the scale
invariant property for $\rho > \rho_c$ and that the dynamic
transition takes place at $\rho_c\simeq 0.6$.

In summary, we have considered the coupled dynamics of a network and a
particle system. In particular, we have considered the rewiring dynamic of a
network which coevolves with diffusing particles.
Our study reveals that the feedback between dynamics and
structure can give rise to a dynamic instability toward the formation of
hubs. This may be one of the origins for the broad degree distribution
observed in real world networks. We have presented the analytic theory
explaining the mechanism for the instability. We have also presented the
scaling theory with which one can understand the dynamic scaling behaviors
and the dynamic phase transition of the model.

Acknowledgement: This work was supported by the Korea Science and
Engineering Foundation (KOSEF) grant funded by Korea government (MOST)
(Grant No. R01-2007-000-10910-0).


\begin{thebibliography}{50}
\bibitem{comment0} One may adopt an alternative protocol where the weight
of an edge increases only when a particle moves through it. We found that this
dynamics does not show an interesting feature.
\bibitem{comment1} Since the weight increment is determined by the degree of
nodes at both ends of an edge, the form $\Delta w_e = \lambda ( K(t) +
\langle k \rangle )$ would be more realistic. We, however, ignore the
constant term for simplicity because it does not modify the conclusion.
\end{thebibliography}
\end{document}